\documentclass[twocolumn, preprintnumbers,amsmath,amssymb]{revtex4}
\usepackage{graphicx}% Include figure files
\usepackage{dcolumn}% Align table columns on decimal point
\usepackage{bm}% bold math
\begin{document}
\title{Lateral organization in mixed lipid bilayers supported on a geometrically patterned substrate}
\author{Qing Liang and Yu-qiang Ma}
\altaffiliation[ ]{Author to whom correspondence should be
addressed. Electronic mail: myqiang@nju.edu.cn.}
\address{National Laboratory of Solid State Microstructures, Nanjing University, Nanjing 210093,
China}

\begin{abstract}
The organization of lipids in biological membranes is essential
for cellular functions such as signal transduction and membrane
trafficking. A major challenge is how to control lateral lipid
composition in supported membranes which are crucial for the
design of biosensors and investigation of cellular processes.
Here, we undertake the first theoretical study of lateral
organization of mixed lipids in  bilayers induced by a
geometrically patterned substrate, and examine the physical
mechanism of patterned substrate-induced structural formation in
the supported lipid bilayers. A rich variety of composition
segregations of lipids are regulated, and the results can account
well for most recent experimental works [Yoon, T. Y.  et al.
 \textit{Nat. Mater.} \textbf{5}, 281(2006) and Parthasarathy, R.
et al.  \textit{Langmuir}  \textbf{22}, 5095(2006)].   The present
study provides a comprehensive understanding of mechanically
controlling the spatial organization of membrane components by
unifying these experimental evidences.

\end{abstract}
\maketitle
\section*{Introduction}
 Biomembranes are self-assembled bilayers of lipid
molecules.  Recently, supported membrane\cite{sackmann1,sackmann2}
has attracted extensive interest, not only due to its importance in
studying the properties and functions of biological membranes, but
also due to its potential applications in design and fabrication of
biological devices.\cite{bayley,cornell,Anrather,Kasemo,figeys}
Supported bilayers can maintain the structural and dynamic
properties of free bilayer such as the lateral fluidity, share many
similarities with the natural membranes, and therefore are widely
used as cell-surface models that connect biological and artificial
materials.\cite{sackmann1,sackmann2} To realize the bio-functional
supported membranes or design novelly biological devices, we usually
need to manipulate the supported membranes in nanometer or
micrometer scale.\cite{Kasemo} For example, one can apply a lateral
electric field to rearrange the distribution of lipids in a
supported membrane for some related cellular
processes.\cite{groves3} Additionally, one can also separate a
supported membrane into well-defined membrane domains by introducing
some barriers into it, and micropattern it by the help of approaches
of photolithographic patterning, microprinting, or microfluidic flow
patterning.\cite{groves1,groves4} Supported membrane micropatterning
has attracted tremendous attention in recent years because of its
potential applications in the investigation of immunobiology, drug
discovery, design of biosensors, etc.\cite{sackmann2}

Most recently, Lee and co-authors\cite{yoon,yoon1,yoon2} presented
an experimental approach to manipulate the lipid segregation in a
mixed supported membrane by use of the geometrical property of
substrate.  They studied the lateral organization of lipids in a
mixed membrane supported on a substrate with a
groove,\cite{yoon,yoon1} and found that, due to
 different effective molecular shapes of mixed lipids, the two kinds of
lipids spontaneously segregate, where the lipids with big
spontaneous curvature are preferentially localized at  the
grooves. In a most recent experimental work, Yoon et
al.\cite{yoon2} studied the effect of   topographical   substrate
on the lipid raft formation in the supported membrane. In their
  system, the topography of the substrate is either
nanocorrugated or nanosmooth, and the membrane supported on such a
substrate will be curved at certain predetermined positions to
follow the substrate structure. Because of   large  bending
rigidity of  sphingolipid/cholesterol-rich $\text{l}_\text{o}$
domains, there exists   a large free energy when they stay in the
curved regions. Thus, the macroscopic rafts which are coarsened by
the nanorafts, can emerge in the nanosmooth regions of substrate,
whereas  there are only nanorafts in the nanocorrugated regions.
On the other hand, Parthasarathy et al.\cite{groves2006} proposed
a curvature-mediated modulation of phase-separated structures in
membranes, and they found that, beyond a critical curvature of
membrane geometry, the cholesterol-rich lipid-ordered
$\text{l}_\text{o}$ domains stayed in the small-curvature regions,
while the cholesterol-poor lipid-disordered $\text{l}_\text{d}$
domains were preferentially localized in the large-curvature
regions due to their different rigidities. Besides the lipid
domains in supported membrane systems, it was also found that
membrane curvature can induce phase separation in the mixed giant
unilamellar vesicles(GUVs) due to the different rigidities of
different domains.\cite{www,roux, bacia}

 Further research will be
needed to uncover and reveal all the possible lipid structures and
physical mechanisms behind experimental evidence from lateral
organization in supported membranes. As is well-known,   it is still
difficult to systematically probe and visualize the laterally
heterogeneous structures  of lipids from tens to hundreds of
nanometers in size with current experimental
technologies.\cite{kraft} There is therefore an urgent need to gain
greater theoretical insight into the physical picture of how
membrane organization is governed. Here, we systematically
investigate the lateral organization  in a mixed bilayer consisting
of symmetric lamella-forming lipids and asymmetric cylinder-forming
lipids supported on a geometrically patterned substrate, and explore
physical mechanism behind structural formation of lipids by using
self-consistent field theory(SCFT). We   expect the present study to
offer a universal and thorough strategy towards efficiently
controlling the lateral phase separation of mixed lipids in bilayers
by varying the substrate roughness, in addition to yielding insights
into several experimental findings.

\section*{Model and Methods}

We consider a system composed of $n_1$   lipid species A and $n_2$
  lipid species B dispersed in $n_s$ hydrophilic homopolymer solvents,\cite{aaa}
which is supported on a geometrically patterned substrate(see Fig.
1).  The headgroup volumes of lipid species A and B  are $v_{h1}$
and $v_{h2}$($v_{h1}<v_{h2})$, respectively, and their tails have
the same length  consisting of $N_t$ segments of segment volume
$\rho_0^{-1}$.\cite{bbb} The solvent chain consists of $N_s$
segments with   segment volume $\rho_0^{-1}$. In the present
problem, we choose $N_t=N_s=N$ without loss of generality. The
conformation of the $\alpha$th chain of lipid A, lipid B, and
homopolymer solvent can be characterized by continuous space
curves ${\bf r}_{\alpha,1} (s)$, ${\bf r}_{\alpha,2}(s)$, and
${\bf r}_{\alpha,s}(s)$, respectively, and all of the chains are
completely flexible.\cite{gh}  We assume that the system is
translational invariant in $y$-direction,\cite{groves2006} then
the surface of patterned substrate is characterized by a
periodical function $S(x)$, which is only dependent on $x$. In our
calculation, we choose $S(x)$ as follows,
%\begin{widetext}
\begin{equation}
S(x)=\begin{cases} H_0,&\text{if}\quad n<\frac{x}{L_0}\leqslant n+\frac{1}{2} ,\\
H_0(1+\sin(\frac{2\pi x}{L_0})),&\text{if}\quad
n+\frac{1}{2}<\frac{x}{L_0}\leqslant n+1 .
\end{cases}
\end{equation}
%\end{widetext}
Here, $H_0$ is the depth of the grooves, $L_0$ is the period of
the substrate, and $n(=0,1,2,\cdots)$ is the number of the periods
 in the patterned substrate.

\begin{figure}
\includegraphics[width=8.cm]{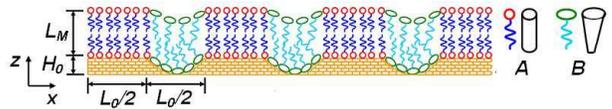}
\caption{ A schematic   of lateral organization  in two-component
lipid bilayer supported on a geometrically patterned substrate.
The lipid species A and B have different effective shapes
(cylindrical- and cone-like) due to different head-tail
symmetries. The substrate is periodically patterned, and each
period is composed of a flat step and an arc-like groove with a
certain curvature. The thickness of the lipid bilayer on the flat
steps is $L_M$, the depth of the grooves is $H_0$, and the
substrate period  is $L_0$. }
\end{figure}

To characterize the system, we define the concentration operators
of headgroups $\hat{\Phi}_{hi}(\mathbf{r})$ and tails
$\hat{\Phi}_{ti}(\mathbf{r})$ of $i$-type lipids($i=1, 2$ denotes
lipid species A and  B, respectively)  and   solvents
$\hat{\Phi}_s(\mathbf{r})$  as follows,
\begin{eqnarray}
&&\hat{\Phi}_{hi}(\mathbf{r})=\gamma_{i}\hat{\rho}_{hi}(\mathbf{r})\;,\\
&&\hat{\Phi}_{ti}(\mathbf{r})=\frac{N}{\rho_0}\sum^{n_i}_{\alpha
=1}\int^{1}_{0}ds\delta(\mathbf{r}-\mathbf{r}_{\alpha,i}(s))\;,
\end{eqnarray}
\begin{eqnarray}
&&\hat{\Phi}_s(\mathbf{r})=\frac{N}{\rho_0}\sum^{n_s}_{\alpha
=1}\int^{1}_{0}ds\delta(\mathbf{r}-\mathbf{r}_{\alpha,s}(s))\;,
\end{eqnarray}
where $\hat{\rho}_{hi}(\mathbf{r})
 =\frac{N}{\rho_0}\sum^{n_i}_{\alpha
=1}\delta(\mathbf{r}-\mathbf{r}_{\alpha,i}(1))$ is the
dimensionless number density of headgroups of $i$-type lipids, and
$\gamma_{i}=v_{hi}\rho_0/N$.

We assume that the system satisfies  the incompressible condition
$\hat{\Phi}_{h1}(\mathbf{r})+\hat{\Phi}_{t1}(\mathbf{r})+\hat{\Phi}_{h2}(\mathbf{r})
+\hat{\Phi}_{t2}(\mathbf{r})+\hat{\Phi}_s(\mathbf{r})=\Phi_0(\mathbf{r})$,
and  $\Phi_0(\mathbf{r})$ is given by,\cite{Matsen1}
\begin{equation}
\Phi_0(\mathbf{r})=\begin{cases} 0, & \text{if}\quad 0\leqslant z<S(x),\\
\frac{1-\cos[\pi(z-S(x))/\epsilon]}{2}, & \text{if}\quad S(x)\leqslant z<S(x)+\epsilon,\\
1, & \text{if}\quad z\geqslant S(x)+\epsilon,
\end{cases}
\end{equation}
where $\epsilon$ is a sufficiently small quantity.

The SCFT method is very powerful to inhomogeneous macromolecular
systems (see, for example, reviews\cite{Schmid,ma33,ma22}) and has
been applied to investigate the lipid bilayer or lipid bulk phase
behaviors in recent years.\cite{leer2,leer5,Li-Schick,
schick,schick5} Leermakers and co-authors used a lattice version of
SCFT to study the lipid-bilayer system,\cite{leer2,leer5} where the
conformation of lipid chains is described by random-walk and RIS
(Rotational Isomeric State) statistics. However, in most of their
works, they assumed that the lipid bilayer was laterally homogeneous
for computational convenience. On the contrary, Schick and
co-authors adopted the off-lattice version of SCFT to study the
lipid systems,\cite{Li-Schick,schick,schick5} where the amphiphilic
nature and chain stretching of lipids are still captured. Compared
to the lattice version of SCFT used by Leermakers and co-workers,
the off-lattice SCFT can account reasonably well for lateral
inhomogeneity of the lipid bilayers, although less molecular details
are considered. Here, we emphasis on the description of laterally
heterogeneous structures and phase separation of lipids due to the
interplay among the conformational entropy of lipids,  effective
molecular shapes of
  lipids, and substrate geometry.  Therefore, the molecular
details of lipids are not important, and it will be more
reasonable to use the coarse-grained model in the framework of
off-lattice SCFT \cite{schick5}.

In the framework of self-consistent field theory(SCFT), we first
transform a hamiltonian of the lipid system into a coarse-grained
field theory description by employing the simple Gaussian model
for lipid chains and treating their interactions through
mean-field Flory-Huggins form.  The partition function of the
system is given by
\begin{equation}
\begin{split}
\mathcal{Z}&=\frac{1}{n_1!n_2!n_s!}\int\prod^{n_1}_{\alpha=1}\hat{\mathcal{D}}\mathbf{r}_{\alpha,1}\prod^{n_2}_{\beta=1}\hat{\mathcal{D}}\mathbf{r}_{\beta,2}
\prod^{n_s}_{\theta=1}\hat{\mathcal{D}}\mathbf{r}_{\theta,s}\\
&\times\delta(\Phi_0(\mathbf{r})-\hat{\Phi}_{h1}(\mathbf{r})-\hat{\Phi}_{t1}(\mathbf{r})-\hat{\Phi}_{h2}(\mathbf{r})-\hat{\Phi}_{t2}(\mathbf{r})\\
&-\hat{\Phi}_s(\mathbf{r}))\times\exp\{-\rho_0\int d\mathbf{r}[\chi
(\hat{\Phi}_{h1}(\mathbf{r})+\hat{\Phi}_{h2}(\mathbf{r})\\
&+\hat{\Phi}_s(\mathbf{r}))(\hat{\Phi}_{t1}(\mathbf{r})+\hat{\Phi}_{t2}(\mathbf{r}))+\chi_{12}(\hat{\Phi}_{h1}(\mathbf{r})\hat{\Phi}_{h2}(\mathbf{r})\\
&+\hat{\Phi}_{t1}(\mathbf{r})\hat{\Phi}_{t2}(\mathbf{r}))-H(\mathbf{r})(\hat{\Phi}_{h1}(\mathbf{r})+\hat{\Phi}_{h2}(\mathbf{r})\\
&+\hat{\Phi}_s(\mathbf{r})-\hat{\Phi}_{t1}(\mathbf{r})-\hat{\Phi}_{t2}(\mathbf{r}))]\}\;,
\end{split}
\end{equation}
%\end{widetext}
where  $ \int \hat{\mathcal{D}}{\mathbf{r}_{\alpha,i} (s)}=\int
\mathcal{D}{\mathbf{r}_{\alpha,i} (s)
\mathcal{P}[\mathbf{r}_{\alpha,i},0,1]}$ is a weighted functional
integral over all   the possible configurations of the $\alpha$th
chain of    lipids or solvent, and
$\mathcal{P}[\mathbf{r}_{\alpha,i},0,1]$  is the probability
distributions of molecular conformations for the $\alpha$th chain.
$\chi$ is the Flory-Huggins interaction parameter between the
hydrophilic components( lipid headgroups or solvents) and the
hydrophobic  lipid tails, and $\chi_{12}$ is the head-head or
tail-tail pair interactions of different kinds of lipids.
$H(\mathbf{r})$ is the surface field of the substrate, and we
set\cite{Balazs1}
\begin{equation}
H(\mathbf{r})=\begin{cases}\infty, & \text{if}\quad 0\leqslant z<S(x),\\
\frac{\Lambda\{1+\cos[\pi(z-S(x))/\epsilon]\}}{\epsilon}, &
\text{if}\quad
S(x)\leqslant z< S(x)+\epsilon,\\
0, & \text{if}\quad z\geqslant S(x)+\epsilon,
\end{cases}
\end{equation}
where $\Lambda$ describes the field strength at the substrate
surface.

The mean-field free energy of the system is given by,\cite{ma33}
\begin{equation}
\begin{split}
&\frac{NF}{\rho_0k_BTV}=-\phi_1f_1\ln(\frac{\Omega_1}{\phi_1f_1V})-\phi_2f_2\ln(\frac{\Omega_2}{\phi_2f_2V})\\
&\quad-(1-\phi_1-\phi_2)\ln(\frac{\Omega_s}{(1-\phi_1-\phi_2)V})\\
&\quad+\frac{1}{V}\int d\mathbf{r}[\chi N
(\varphi_{h1}(\mathbf{r})+\varphi_{h2}(\mathbf{r})+\varphi_s(\mathbf{r}))(\varphi_{t1}(\mathbf{r})\\
&\quad+\varphi_{t2}(\mathbf{r}))
+\chi_{12}N(\varphi_{h1}(\mathbf{r})\varphi_{h2}(\mathbf{r})+\varphi_{t1}(\mathbf{r})\varphi_{t2}(\mathbf{r}))\\
&\quad-H(\mathbf{r})N(\varphi_{h1}(\mathbf{r})+\varphi_{h2}(\mathbf{r})
+\varphi_s(\mathbf{r})-\varphi_{t1}(\mathbf{r})\\
&\quad-\varphi_{t2}(\mathbf{r}))
-w_{h1}(\mathbf{r})\rho_{h1}(\mathbf{r})-w_{t1}(\mathbf{r})\varphi_{t1}(\mathbf{r})\\
&\quad-w_{h2}(\mathbf{r})\rho_{h2}(\mathbf{r})-w_{t2}(\mathbf{r})\varphi_{t2}(\mathbf{r})-w_s(\mathbf{r})\varphi_s(\mathbf{r})\\
&\quad-\xi(\mathbf{r})(\Phi_0(\mathbf{r})-\varphi_{h1}(\mathbf{r})-\varphi_{t1}(\mathbf{r})-\varphi_{h2}(\mathbf{r})\\
&\quad-\varphi_{t2}(\mathbf{r})-\varphi_s(\mathbf{r}))],
\end{split}
\end{equation}
 where  $k_B$ is Boltzmann's constant, $T$ is the temperature, and
  $V=\int \Phi_0(\mathbf{r})
d\mathbf{r}$ is the effective volume of the system.\cite{ma22}
$\phi_i=n_i(v_{hi}+N\rho^{-1}_0)/V$ and
$f_i=N\rho^{-1}_0/(v_{hi}+N\rho^{-1}_0)$ are the overall volume
fraction and the tail volume fraction of lipid species A ($i=1$)
or B ($i=2$) in the system, respectively. $\Omega_1$, $\Omega_2$,
and $\Omega_s$ are the single-molecule partition functions  of
lipid A, lipid B, and solvent, respectively, and they are given by
$\Omega_i=\int\hat{\mathcal{D}}\mathbf{r}_{i}(s)
\exp[-w_{hi}(\mathbf{r}_{i}(1))
 -\int^1_0dsw_{ti}(\mathbf{r}_{i}(s))]$ ($i=1, 2$)  and
$\Omega_s=\int\hat{\mathcal{D}}\mathbf{r}_{s}(s)
\exp[-\int^1_0dsw_s(\mathbf{r}_{s}(s))]$. The variables
$\varphi_{hi}$, $\varphi_{ti}$, and $\varphi_s$ are the
concentration functions of $i$-type headgroups, $i$-type tails, and
solvents, while $w_{hi}$, $w_{ti}$, and $w_s$ are the fields acting
on the headgroups, tail segments, and solvent segments,
respectively. $\xi(\mathbf{r})$ is a Lagrange-multiplier field which
ensures the incompressibility of the system. The  fields and
densities  are then obtained by minimizing  the free energy in
Eq.(8), and given as
\begin{eqnarray}
&& w_{h1}(\mathbf{r})=\chi
N\gamma_1(\varphi_{t1}(\mathbf{r})+\varphi_{t2}(\mathbf{r}))\nonumber\\
&&\qquad\qquad+\chi_{12}N\gamma_1\varphi_{h2}(\mathbf{r})-\gamma_1H(\mathbf{r})N+\gamma_1\xi(\mathbf{r}),\\
&& w_{t1}(\mathbf{r})=\chi
N(\varphi_{h1}(\mathbf{r})+\varphi_{h2}(\mathbf{r})+\varphi_s(\mathbf{r}))\nonumber\\
&&\qquad\qquad+\chi_{12}N\varphi_{t2}(\mathbf{r})+H(\mathbf{r})N+\xi(\mathbf{r}),
\end{eqnarray}
\begin{eqnarray}
&& w_{h2}(\mathbf{r})=\chi
N\gamma_2(\varphi_{t1}(\mathbf{r})+\varphi_{t2}(\mathbf{r}))\nonumber\\
&&\qquad\qquad+\chi_{12}N\gamma_2\varphi_{h1}(\mathbf{r})-\gamma_2H(\mathbf{r})N+\gamma_2\xi(\mathbf{r}),\\
&& w_{t2}(\mathbf{r})=\chi
N(\varphi_{h1}(\mathbf{r})+\varphi_{h2}(\mathbf{r})+\varphi_s(\mathbf{r}))\nonumber\\
&&\qquad\qquad+\chi_{12}N\varphi_{t1}(\mathbf{r})+H(\mathbf{r})N+\xi(\mathbf{r}),\\
&& w_{s}(\mathbf{r})=\chi
N(\varphi_{t1}(\mathbf{r})+\varphi_{t2}(\mathbf{r}))
-H(\mathbf{r})N+\xi(\mathbf{r}),\\
&&\rho_{h1}(\mathbf{r})=-\frac{\phi_1f_1V}{\Omega_1}\frac{\partial\Omega_1}{\partial
w_{h1}(\mathbf{r})}\nonumber\\
&&\qquad\quad=\frac{\phi_1f_1V}{\Omega_1}q_1(\mathbf{r},1)q^{\dag}_1(\mathbf{r},1),\\
&&\varphi_{h1}(\mathbf{r})=\gamma_1\rho_{h1}(\mathbf{r}),\\
&&\varphi_{t1}(\mathbf{r})=-\frac{\phi_1f_1V}{\Omega_1}\frac{\partial
\Omega_1}{\partial
w_{t1}(\mathbf{r})}\nonumber\\
&&\qquad\quad=\frac{\phi_1f_1V}{\Omega_1}\int^1_0 d s
q_1(\mathbf{r},s)q^{\dag}_1(\mathbf{r},s),
\end{eqnarray}
\begin{eqnarray}
&&\rho_{h2}(\mathbf{r})=-\frac{\phi_2f_2V}{\Omega_2}\frac{\partial\Omega_2}{\partial
w_{h2}(\mathbf{r})}\nonumber\\
&&\qquad\quad=\frac{\phi_2f_2V}{\Omega_2}q_2(\mathbf{r},1)q^{\dag}_2(\mathbf{r},1),\\
&&\varphi_{h2}(\mathbf{r})=\gamma_2\rho_{h2}(\mathbf{r}),\\
&&\varphi_{t2}(\mathbf{r})=-\frac{\phi_2f_2V}{\Omega_2}\frac{\partial
\Omega_2}{\partial
w_{t2}(\mathbf{r})}\nonumber\\
&&\qquad\quad=\frac{\phi_2f_2V}{\Omega_2}\int^1_0 d s
q_2(\mathbf{r},s)q^{\dag}_2(\mathbf{r},s),\\
&&\varphi_s(\mathbf{r})=-\frac{(1-\phi_1-\phi_2)V}{\Omega_s}\frac{\partial
\Omega_s}{\partial w_s(\mathbf{r})}\nonumber\\
&&\thickspace\thickspace\qquad=\frac{(1-\phi_1-\phi_2)V}{\Omega_s}\int^1_0d
s
q_s(\mathbf{r},s)q_s(\mathbf{r},1-s),\\
&&\sum^{2}_{i=1}(\varphi_{hi}(\mathbf{r})+\varphi_{ti}(\mathbf{r}))+\varphi_s(\mathbf{r})=\Phi_0(\mathbf{r}),
\end{eqnarray}
where $q_i(\mathbf{r},s)$ and $q^\dag_i(\mathbf{r},s)(i=1,2)$ are
the end-segment distribution functions of the $i$-type lipids, and
are defined by,\cite{ma33}
\begin{eqnarray}
&&q_i(\mathbf{r},s)=\int
\mathcal{D}\mathbf{r}_{\alpha,i}\mathcal{P}[\mathbf{r}_{\alpha,i},0,s]\delta(\mathbf{r}-\mathbf{r}_{\alpha,i}(s))\nonumber\\
&&\quad\quad\exp{\{-\int^s_0dtw_{ti}(\mathbf{r}_{\alpha,i}(t))\}},\\
&&q^{\dag}_i(\mathbf{r},s)=\int
\mathcal{D}\mathbf{r}_{\alpha,i}\mathcal{P}[\mathbf{r}_{\alpha,i},s,1]\delta(\mathbf{r}-\mathbf{r}_{\alpha,i}(s))\nonumber\\
&&\quad\quad\exp{\{-w_{hi}(\mathbf{r}_{\alpha,i}(1))-\int^1_sdtw_{ti}(\mathbf{r}_{\alpha,i}(t))\}}.
\end{eqnarray}
They satisfy the modified diffusion equations,
\begin{eqnarray}
&&\frac{\partial q_i}{\partial s}=\frac{Na^2}{6}\nabla^2q_i-w_tq_i\;,\\
&&\frac{\partial q^{\dag}_i}{\partial
s}=-\frac{Na^2}{6}\nabla^2q^{\dag}_i+w_tq^{\dag}_i \;,
\end{eqnarray}
where $a$ is the statistical segment length. The initial conditions
are $q_i(\mathbf{r},0)=1$
 and $q^{\dag}_i(\mathbf{r},1)=\exp(-w_{hi}(\mathbf{r}))$.
$q_s(\mathbf{r},s)$     has the analogous definition  for the
solvent, which satisfies  Eq. (24) with the initial condition
$q_s(\mathbf{r},0)=1$.  The resulting self-consistent equations (9)
- (21) together with equations(22)-(25) can be numerically solved by
the combinatorial screening method proposed by Drolet and
Fredrickson.\cite{ma22}

We choose  the tail volume fractions in lipid species A and B as
$f_1=0.6$ and $f_2=0.45$,    and the other parameters are fixed to
be $\chi N=20$, $N=50$, $a=1$, and $\rho^{-1}_0=1$. Thus, we have
$\gamma_1=0.67$, $\gamma_2=1.22$, $v_{h1}=33$, and $v_{h2}=61$, and
with these parameters, lipid species A and B form a lamellar and a
hexagonally cylindrical phases in bulk,
respectively.\cite{Li-Schick} This means that lipid species A has a
symmetric head/tail shape favoring the packing of flat interfaces,
while lipid species B is asymmetric, which prefers to curve the
interfaces to the lipid tails. As  lipid species A and B are mixed
to put onto a hydrophilic substrate, a mixed lipid bilayer may be
formed. Here we take the overall volume fractions of lipid species A
and B as $\phi_1=\phi_2=0.12$, and assume that the repulsive
head-head/tail-tail  interactions between the two kinds of lipids
are weak($\chi_{12}N=3$). The strength of  surface field is taken to
be $\Lambda N=1$. To facilitate the formation of a bilayer
structure, we adopt the scenario proposed in Ref. 29 by introducing
an external field favoring the lipid tails in the initial stage of
the self-consistent calculation, which enforces the bilayer membrane
formed along the substrate surface profile. When the bilayer is
formed, the external field is turned off and the calculation
continues until the equilibrium state of  system is reached.

\section*{Results and Discussion}
\begin{figure*}
\includegraphics[width=15cm]{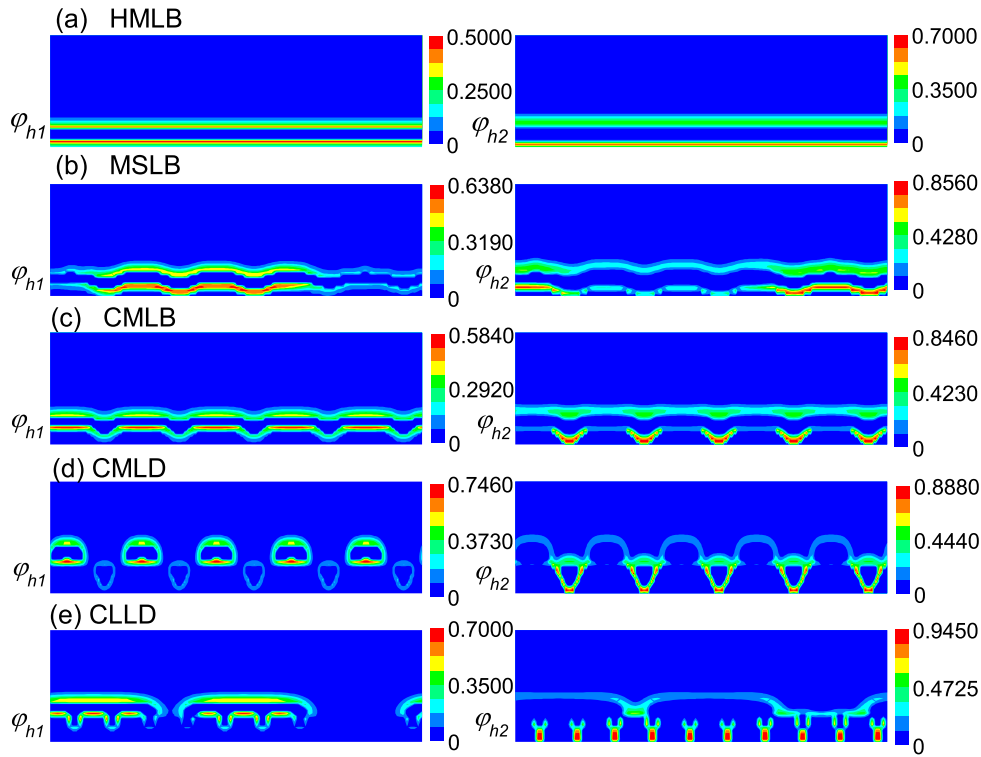}
\caption{Summary of all observed morphologies for headgroup of
lipid spcies A ($\varphi_{h1}$) and
 of lipid species B ($\varphi_{h2}$): (a)
homogeneously  mixed lipid bilayer(HMLB), (b) macro-segregated
lipid bilayer(MSLB), (c) component-modulated lipid bilayer(CMLB),
(d) component-modulated lipid domain(CMLD),  and (e) coalesced
long lipid droplet(CLLD) structures.}
\end{figure*}

\begin{figure}
\includegraphics[width=6.cm]{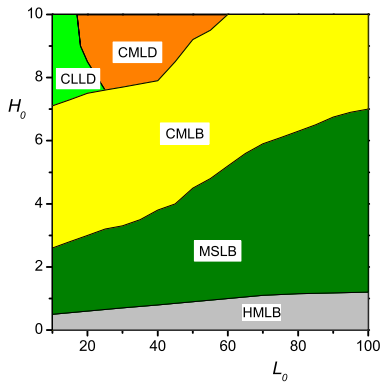}
\caption{ Morphology diagram for the mixed lipid bilayer supported
on a geometrically patterned substrate.}
\end{figure}
We now examine the formation of all the possible lipid structures
by systematically varying the groove depth ($H_0$) and the
substrate period ($L_0$),  and the results are summarized  in Fig.
2.  Figure 3 shows a morphology diagram, and  a rich variety of
lipid domains can be tuned by varying $L_0$ and $H_0$. Here, the
height of system is fixed to be 40 which is large enough to not
influence the morphology of lipid bilayer, and the lateral length
ranging from $100 \sim 500$ is adjusted to search the stable
phases by the total free energy minimization. The substrate period
changes from $L_0=10$ to $L_0=100$ and   the groove depth from
$H_0=0$ to $H_0=10$, which are either comparable to or several
times larger than the natural size of the
lipids($R_0=a\sqrt{N}\simeq 7$), and therefore, the evolved scale
of lateral organization is within the nanometer size(i.e., from
tens to hundreds of nanometers).

When $H_0$ is vanishingly small,  the perturbation from the
substrate geometry is neglectable, and  lipids are mixed uniformly
on the substrate due to mixing entropy. Thus  a homogeneously
mixed lipid bilayer(HMLB) state is formed(Fig. 2(a)). With the
increase of $H_0$, the substrate becomes slightly rough, and thus
lipid species B will first occupy the grooves to gain more
conformational entropy. This is because that, although the
substrate is relatively flat, there still exist some perturbations
to the lipid bilayer from the substrate where the grooves drive
the formation of lipid B-rich islands(nucleation). As lipid B-rich
islands initiates from the grooves, a lateral expansion of the
islands on the substrate appears, and eventually leads to the
formation of macro-segregated lipid bilayer(MSLB) under the help
of the lipid A - B interfacial energy (Fig. 2(b)). When the groove
depth $H_0$ is increased to modest values  where the curvature of
grooves becomes comparable to the bulk curvature of lipid species
B,   both lipid species B and A can get more entropy as they stay
in the grooves and on the flat steps, respectively. Thus, a
component-modulated lipid bilayer(CMLB) structure appears in this
region, as shown in Fig. 2(c).

When $H_0$ is enough large, the groove curvature may be larger
than the spontaneous  curvature of lipid species B at small or
modest $L_0$ region, and the  lipid distribution  may become  more
complicated. In this case, if the bilayer structure is still
maintained, the bilayer would be curved strongly in the regime of
grooves, and thus lipid species B in the upper leaflet must lose
plenty of conformational entropy because the curvature of the
upper leaflet is in contrast to their bulk curvature. However,
lipid species B in the lower leaflet prefers to fill in the
grooves, compared to the case if   lipid species A fills  in the
grooves. As a result, the lipid bilayer may be ruptured by the
extremely rough substrate: the component-modulated lipid
domains(CMLD)   appears at modest $L_0$, and the coalesced long
lipid droplets(CLLD)   with spreading lipid-A droplets is formed
at smaller $L_0$ (see Fig. 3). The formation of the CMLD and CLLD
structures in Fig. 2 can be understood by the help of the
morphologies of thin films on chemically patterned substrates.
Actually, in the CMLD   and   CLLD structures, the lipid species B
almost  fill up the grooves, and therefore for lipid species A,
the substrate now becomes a chemically patterned one which is
composed of alternating more-wettable hard-wall parts and
less-wettable lipid-B parts. Thus, the problem can be highlighted
by templating of thin films on chemically patterned substrates
studied by Kargupta and Sharma\cite{kajari}. They examined the
effects of the periodicity of substrate pattern and the width of
the more-wettable stripes on the morphology of thin films, and
found that when the patterned substrate periodicity  is beyond a
critical length, the thin film pattern closely replicates the
substrate surface pattern, namely dewetting occurs on every
less-wettable parts of the patterned substrate, just as shown in
Fig. 2(d) where  lipid species A  is almost localized on
more-wettable hard-surface parts. On the contrary, when the
periodicity is smaller than the critical length, the dewetting
effect may partially  be suppressed, and the lone droplet of thin
films spans across the remaining less-wettable sites, exactly like
the morphology of  lipid-A droplets shown in Fig. 2(e).
 In real biological systems, it was reported \cite{sara} that
 large quantities of membrane fusion pore
sites which are concentrated in cone-shaped  nonlamellar lipids
such as phosphatidylethanolamine(PE), are simultaneously formed at
high-curvature regions. This provides further evidence
 implying the validity of these  unexpected structures
in Figs. 2(d) and 2(e). Therefore, the lipid heterogeneity by
forming CLLD and CMLD structures may highlight the understanding
of the relation between the membrane curvature and fusion,
fission, or pore formation of membranes. In the large-$L_0$
region, the curvature of the groove become comparable to the
spontaneous curvature of lipid species B again, and a CMLB phase
is presented there.

Importantly, by changing the position of grooves for   given
$H_{0}$ and $L_0$, we find that at arbitrary spatial distribution
of grooves, the above conclusions keep unchanged, independent of
distributed positions of grooves. Actually, such a substrate
geometry mediated phase-separated behavior   of lipids provides a
controllable mechanism for spatial sorting of membrane components
by varying the roughness of substrates. Furthermore, this is also
helpful to understanding how cellular membrane geometry to govern
the lipid ordering and domain rafts in real biological systems.

\begin{figure}
\includegraphics[width=9cm]{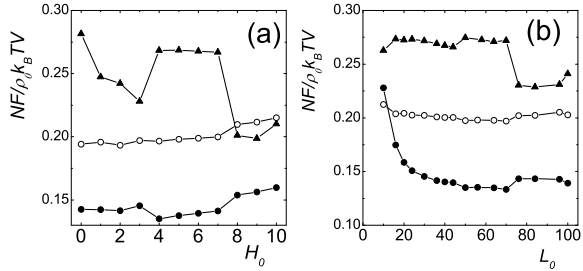}
\caption{Free energy   contributions from the entropy of   lipid
species A ($\circ$), the entropy of
  lipid species B ($\bullet$), and interfacial energy between
the two kinds of lipids($\blacktriangle$). (a) $L_0=40$; (b)
$H_0=6$.}
\end{figure}

To clarify the underlying mechanism  of formed phase behavior  of
mixed lipids,  we examine  the competition of the interfacial
energy, conformational entropy, and mixing entropy of lipids.
Figure 4(a) shows the variation of free energy $F$ of the system
versus $H_0$ by fixing $L_0 = 40$.  When $H_0=0$ (the substrate is
completely smooth), the mixing entropy dominates the lateral
organization of the two kinds of lipids, leading to the uniform
dispersion of lipids. When $H_0$ is low (e.g., $H_0=1$$\sim$ 3),
the interfacial energy dominates the distribution of  lipids and a
MSLB structure appears with lateral growth of lipid B-islands
located at grooves. For middle values of $H_0$ (e.g.,
$H_0=4$$\sim$ 7), a CMLB structure is formed by releasing the
conformational entropy of lipid species B, while at the same time,
the interfacial energy is increased. Such a CMLB structure is
mainly contributed to the entropy effect of   lipid species B.
When $H_0$ is large enough (e.g., $H_0=8$ $\sim$ 10), the
formation of CMLD structure  where the two kinds of lipids
separate   strongly, largely decreases   the interfacial energy,
while also leads to the loss of mixing entropy. In Fig. 4(b), we
show the variation of free energy versus $L_0$ by fixing $H_0 =
6$. The mixed lipid bilayer is either component-modulated
structure for small  $L_0$   as a result of the release of
conformational  entropy of lipid species B  or a macro-separated
bilayer for large $L_0$   where the interfacial energy is greatly
decreased.

It is vitally useful to make a comparison between the present work
and experimental works.\cite{yoon,yoon1,yoon2,groves2006} First, the
appearance of CMLB structure accurately confirms  the experimental
results observed in Refs. 11 and 12. We also find that such a CMLB
structure formed in the modest curvature regions exactly corresponds
to the nanorafts emerged in the nanocorrugated regions in Ref. 13
and the spatial modulated structure of alternating the
$\text{l}_\text{o}$ domains in low curvature and $\text{l}_\text{d}$
domains in high curvature regions observed in Ref. 14, while the
MSLB structure presented in the lower curvature regions is just as
the case that the macroscopic $\text{l}_\text{o}$ domains appear in
nanosmooth regions in Ref. 13. Further, our phase diagram describing
all the observed morphology regimes, which greatly enriched
 the lateral organization of membrane component on a geometrically patterned
 substrate, will be very helpful to experimentally probe the lipid
distribution on a wide range of length scales of
nanometers.\cite{kraft} Finally,  by comparing the physical models
of the experimental works and our theoretical work, we easily find
that although our system is not completely identical to those
studied by Yoon et al.\cite{yoon2} and Parthasarathy et
al.\cite{groves2006}, the physics of these models is consistent,
namely the curved regions offer some barriers for lipid-A domains in
the present work and $\text{l}_\text{o}$ domains in Refs. 13 and 14.
This is because the sphingolipid/cholesterol-rich
$\text{l}_\text{o}$ domain and cholesterol-poor $\text{l}_\text{d}$
domain  can reasonably be represented by symmetric lipid A and
asymmetric lipid B in our model, respectively. The only difference
during the lipid organization is that the free energy barrier in the
experimental works is provided by the bending energy of lipid
domains, while in our system, comes from the loss of conformational
entropy of lipids.

\section*{Conclusions}
We conducted a thorough investigation in the lateral organization of
a two-component lipid bilayer on a geometrically patterned substrate
by using self-consistent field theory. Because of their different
molecular shapes, the two kinds of lipids prefer to stay in the
grooves and flat steps of substrate, respectively. By adjusting the
groove depth  and the substrate period, we obtained a rich variety
of laterally organized lipid structures.  We also examine the
microscopic mechanism of self-assembly of lipids in the mixed
bilayer supported on the geometrically patterned substrate, and find
that the resulting phase behavior of mixed lipids   is dominated by
the competition of the interfacial energy, conformational entropy,
and mixing entropy of lipids. The present study confirmed recent
experimental results in Refs. 11 and 12,  and qualitatively agreed
with the experimental findings of Refs. 13 and 14. Most importantly,
we also predict, for the first time, the formation of the CLLD and
CMLD structures in large curvature regions, which may provide
theoretical insight into future experiment with the improvement of
visualization techniques towards smaller scales of geometry and help
to understand the microscopic mechanism of fusion pore formation in
biomembranes.

\section*{Acknowledgments} This work was supported by the National Natural
Science Foundation of China, No. 10334020, No. 10021001,  No.
20674037, and No. 10574061.

%\newpage


\begin{references}
\bibitem{aaa} In real system, the solvent is usually
composed of small water molecules. However, in the SCFT framework,
if the solvent is modeled as small featureless molecules, one
finds that the small molecules are very difficult to be removed
from the hydrophobic regions of the lipid bilayers due to large
mixing entropy.\cite{leer5} Actually, the main goal of the present
paper is to reveal the lateral organization of lipid bilayer
supported on a patterned substrate, where the solvent  only offers
a hydrophilic background to maintain the lipid bilayer and could
not affect the physics of lipid organization. This means that the
conformation of solvent is not important, and therefore, we can
use the homopolymer solvents instead of small water molecules, as
 used by Schick and co-authors\cite{schick,schick5}.
\bibitem{bbb} Each lipid consists of one hydrophilic headgroup and one
hydrophobic tail: the head-tail volume ratio  signifies  the
effective molecular shape of  lipids    reflecting the head-head
and tail-tail interactions, and single tail displays the chain
stretching  of lipids. Further, at the coarse-grained level, the
lipids A and B can reasonably represent cholesterol-rich
lipid-ordered ($l_{o}$) domain and cholesterol-poor
lipid-disordered ($l_{d}$) domain\cite{yoon2}, respectively,
although the real lipid is usually composed of one headgroup and
two tails.
\bibitem{gh} Li and Schick \cite{Li-Schick} have compared the SCFT results of
the lipid system with the experimental results, and a good
agreement was found between them, while  the real lipid is not
completely flexible.
\end{references}
\end{document}